# Room-temperature printing of ultrathin Quasi-2D GaN semiconductor via liquid metal gallium surface confined nitridation reaction

Qian Li [1,#], Bang-Deng Du [1,#], Jian-Ye Gao [1,3] Bao-Yu Xing [4], Dian-Kai Wang [4], Ji-Fei Ye [4] & Jing Liu [1,2,*]

1. Technical Institute of Physics and Chemistry, Chinese Academy of Sciences, Beijing 100190, P. R. China.
2. Department of Biomedical Engineering, School of Medicine, Tsinghua University, Beijing 100084, P. R. China
3. School of Engineering Science, University of Chinese Academy of Sciences, Beijing 100190, P. R. China
4. State Key Laboratory of Laser Propulsion and Application, Space Engineering University, Beijing 101416, P. R. China
* Corresponding Author; E-mail addresses: J. Liu (jliu@mail.ipc.ac.cn)
# Q. L. and B. D. D contributed equally to this work.

**Abstract**

Outstanding wide-bandgap semiconductor material such as gallium nitride (GaN) has been extensively utilized in power electronics, radiofrequency amplifiers, and harsh environment devices. Due to its quantum confinement effect in enabling desired deep-ultraviolet emission, excitonic impact, and electronic transport features, two-dimensional (2D) or ultrathin quasi-2D GaN semiconductors have been one of the most remarkable candidates for future growth of microelectronic devices. Here, for the first time, we reported a large area, wide bandgap, and room-temperature quasi-2D GaN synthesis and printing strategy through introducing the plasma medicated liquid metal gallium surface-confined nitridation reaction mechanism. The developed direct fabrication and compositional process is consistent with various electronics manufacturing approaches and thus opens an easy going way for cost-effective growth of the third-generation semiconductor. In particular, the fully printed field-effect transistors relying on the GaN thus made show p-type switching with an on/off ratio greater than $10^5$, maximum field-effect hole mobility of 53 $cm^2$ $V^{-1}$ $s^{-1}$, and a small sub-threshold swing. As it was demonstrated, the present method allows to produce at room temperature the GaN with thickness spanning from 1nm to nanometers. This basic method can be further extended, generalized, and utilized for making various electronic and photoelectronic devices in the coming time.

## 1. Introduction

Semiconductor technology has been one of the most fundamental cores in developing the integrated circuit industry. Since witness of the bottleneck of the first two generations of semiconductors, the third-generation high-temperature wide-band-gap semiconductor nanomaterials such as gallium nitride (GaN)[1], zinc oxide (ZnO), aluminum nitride (AlN)[2], silicon carbide (SiC)[3,4], and diamond etc., have become a focus in recent years. GaN nanomaterials are novel semiconductors owning many outstanding merits like high saturated electron mobility, radiation resistance, acid and alkali corrosion resistance, large thermal conductivity, and strong breakdown field. In the GaN electronic devices, the inevitable electron transfer from the valence band to the conduction band can be suppressed through the wide energy bandgap. The ambient energy from strong electric fields, high temperature, and intense energy particles can activate the mentioned electron transfer[5]. Accordingly, the devices can retain their electrical features in various scenarios. For such reason, the GaN was regarded as one of the most attractive semiconductor materials due to its idealistic efficiency and stability[6,7]. Tremendous efforts have therefore been devoted to developing and analyzing the GaN materials throughout the world[8–10].

Compared with the bulk GaN materials, various nanoscale effects of the low dimensional GaN materials can display better photoelectric, mechanical, thermal stability, as well as electrical[11,12], and chemical features[13]. Apart from the fundamental physicochemical features of GaN, they also own the surface, small-size, and quantum-confinement impacts as one of the most exciting areas for future growth of microelectronic devices. However, until now it is still not easy to construct a low-dimensional GaN. Only rather limited outstanding works are available. Syed et al. proposed a two-step process to synthesis two-dimensional (2D) GaN nanosheets, the process includes obtaining 2D $Ga_2O_3$ by extrusion printing, and then converting $Ga_2O_3$ to GaN through ammonolysis in a tubular furnace[14]. Chen et al. reported the development of 2D GaN single crystals (about 4.1 nm) attained using a surface-confined nitridation reaction (SCNR) through the chemical vapor deposition (CVD)[15]. In 2016, Balushi et al. employed the migration-enhanced encapsulated growth method to construct 2D GaN monolayer nanosheets[16]. Unfortunately, the temperatures used in most of the process are above 500 °C that is inconsistent with various electronic industry operations. This is because the long run time of the deposition process can increase the cost and feasibility. Clearly, the construction and research technology of low-dimensional GaN materials and devices should be developed and enhanced to satisfy the appeals from so many practical needs. In this side, a low temperature preparation of large-area, high-quality and uniform GaN films will generate a significant impact on high thermal stability 2D integrated circuit industry designed for power electronics. However, so far, there is still no report on the room temperature preparation of ultrathin or 2D GaN films.

Herein, we proposed and demonstrated for the first time to print the ultrathin quasi-2D GaN films on $SiO_2$/Si substrates through introducing a liquid metal-based synthesis and printing processes at room temperature. The basic principle relies on using nitrogen plasma to trigger nitriding of gallium droplets at room temperature, and then transferring them to the substrate through the proposed van der Waals (vdW) printing technology. The whole preparation process is rather easy going which is completely performed at room temperature and is consistent with the current electronics manufacturing processes. The proposed method can produce nanometer scaled thickness GaN film attained via individual or multiple prints. Moreover, we report excellent electronic performance of the printed ultrathin quasi-2D GaN. For illustration purpose, fully printed side-gated field-effect transistors (FETs) are fabricated, from which the quasi-2D GaN-FETs exhibit outstanding performance with a large current on/off ratio (> $10^5$), high field-effect mobility (~53 $cm^2\ V^{-1}\ s^{-1}$), and tiny subthreshold slopes (~ 98 $mV\ dec^{-1}$) with a high degree of reproducibility. This study suggests a reliable and straightforward room temperature large-scale manufacturing technique for printing quasi-2D GaN with outstanding working features, which opens significantly practical potential for wafer-scale processes. It also paves the way for the application of GaN semiconductor in a new generation of all printed electronic devices, integrated circuits and more functional devices.

## 2. Results and Discussion

**Liquid metal-based ultrathin quasi-2D GaN printing technology.** Conventionally, quite a few processes have been available to prepare GaN films, such as metal organic chemical vapor deposition (MOCVD) and atomic layer deposition (ALD), which are illustrated in Fig. 1a and b. The high preparation temperature (> 250 °C) and toxic material ($Ga(CH_3)_3$) are main disadvantages of these processes to produce GaN at a large industrial scale. Recently, room temperature $N_2$ plasma treatment as an emerging tool for green nitrogen fixation and surface modification of materials, have been tried in the field of ammonia synthesis[17], N doped in 2D materials[18], plasma enhanced CVD depositing thin film, nitride layer formation on rigid materials, etc. The $N_2$ plasma technology owns merits of low cost, high efficiency and pollution-free on nitriding treatment of material surfaces. With this in mind, here a simple and convenient process of GaN thin film prepared through $N_2$ plasma irradiating clean liquid Ga surface is proposed.

The scheme in Fig. 1c illustrates the mechanism of our introduced process of GaN formation. Such plasma nitriding process employs an unusual glow discharge involving high current and charge densities. A potential difference is generated by applying a DC voltage between the discharger (anode) and the liquid Ga (cathode), which ionizes the nitrogen to produce the glow discharge. Generally, atomic nitrogen ($N_{atom}$), excited nitrogen molecules ($N_2^*$), activated nitrogen molecule ions ($N_2^+$) and electrons ($e^-$) are main contents of nitrogen plasma gas[17]. In the plasma nitriding of gallium, the accelerated $N_{atom}$, $N_2^*$, $N_2^+$ all can hit the surface of the liquid Ga with high kinetic energy, and react directly with Ga atom to form GaN molecule. In a word, producing GaN from intimate reaction of nitrogen and liquid Ga in plasma can be represented as: $N_2 + 2Ga \xrightarrow{plasma} 2GaN$. Because the chemical reaction of nitrogen plasma is a thermodynamically stable excited state and ionic state, and the reaction activation energy is pretty low, it is much easier to generate GaN by the reaction of nitrogen plasma with liquid Ga than that through the traditional nitriding reaction (MOCVD) $Ga(CH_3)_3 + NH_3 \xrightarrow{above\ 950°C} GaN + 3CH_4$.

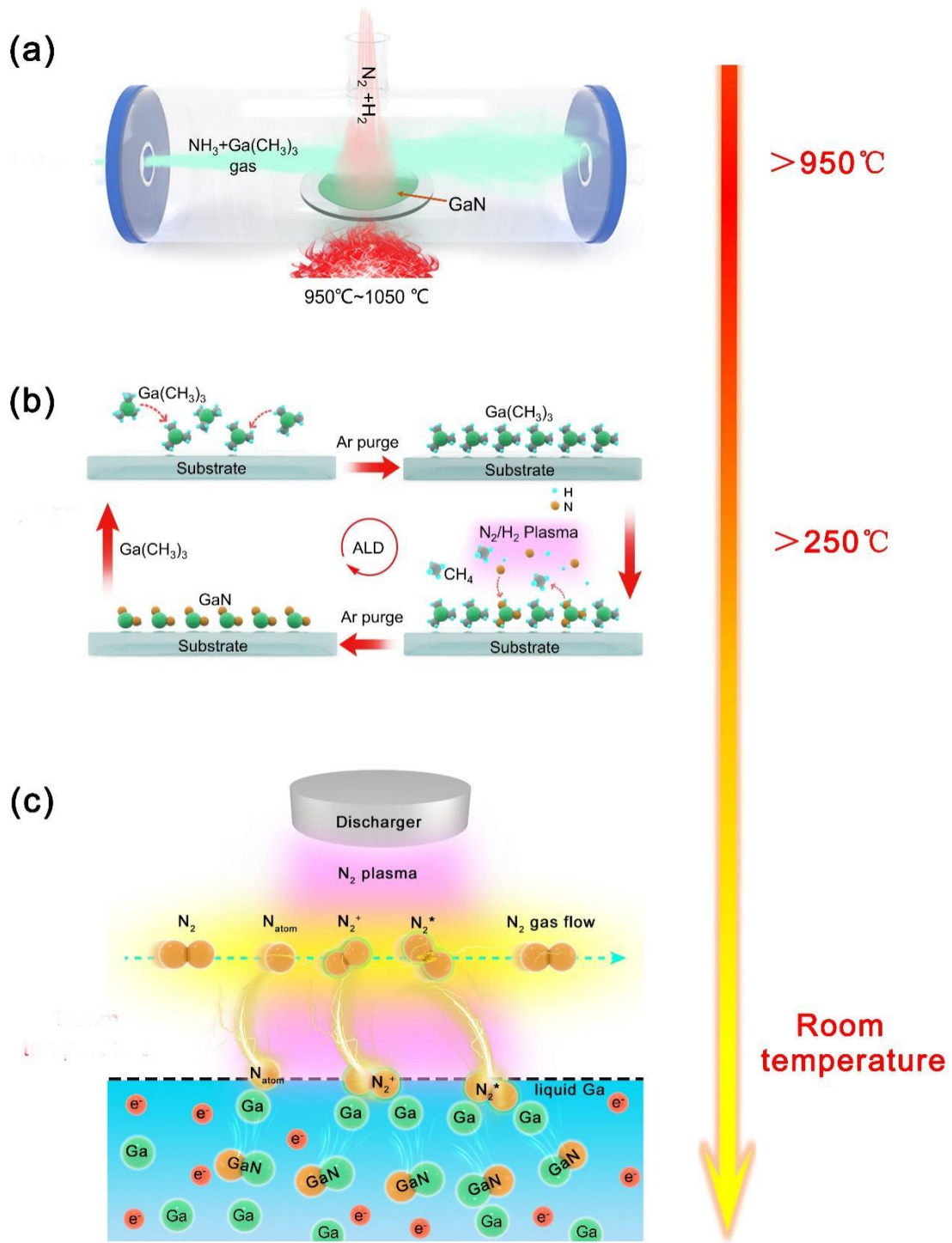

**Fig. 1 The schematic diagram of three GaN growth processes and temperatures.** (a) Illustration of MOCVD locus to fabricate GaN, at temperature above 950 °C. (b) Illustration of ALD locus to prepare GaN, at temperature above 250 °C. (c) Illustration of nitrogen plasma-liquid Ga reaction locus to harvest GaN, at room temperature.

The schematic in Fig. 2a illustrates the preparation of ultrathin quasi-2D GaN on the surface of liquid Ga droplet using $N_2$-plasma treatment technology at room temperature. The changes of fresh Ga droplet before and after nitrogen plasma treatment are shown in Supplementary Fig. 1. The schematic setup of the system to transform the liquid Ga to GaN is elucidated in Fig. 2b. The formation of GaN thin films was realized by nitrogen plasma triggering surface limited nitriding reaction. Gallium droplets with plasma treated surface were placed on SiO$_2$/Si substrate and moved on the surface with the help of scraper, during the scraping process of liquid Ga, the surface nitride produced by plasma bombardment in nitrogen environment sticks to the substrate through van der Waals force and the clean Ga is exposed (Fig. 2c) (Detailed description of the process parameters will be given in the Methods.). The nitridation reaction and printing of Ga both are conducted at room temperature that is consistent with the current electronic device production processes. The covered substrate area can be expanded by choosing a larger droplet diameter and a longer travel distance of the scraper. The thickness of the film can be increased via expanding nitrogen plasma trigger time or frequent printing (Methods and Supplementary Fig. 2). Fig. 2(d) shows one touch print of the GaN film obtained from the surface of Ga droplets after being treated 10 min by nitrogen plasma. It revealed a large and continuous ultrathin GaN film reaching lateral dimensions more than many centimeters. Based on the atomic force microscopy (AFM), the thickness of the deposited GaN layer is measured as ~4 nm that is moderately larger than a single GaN unit cell (Fig. 2e). The printable GaN is slightly thicker than the single layer, indicating that the prepared GaN film is an ultrathin quasi-2D film. AFM also indicates that the printed GaN film's surface roughness is analogous to that of the SiO$_2$ substrate, demonstrating that the GaN film has minimum cracks, holes, folds, or bubbles, reflecting conformal and homogeneous attachment.

It is noteworthy that the thickness of singly printed GaN film is below 1 nm (Supplementary Fig. 2 and Fig. 3a), when the nitrogen plasma trigger time is lower than 1 min. The harvested GaN film can be classified as 2D materials. However, cracks and holes can also be seen on obtained 2D GaN. Actually, the uniformity and continuity of printed GaN films are not always good enough, when the nitrogen plasma trigger time is lower than 5 min. It might be because the insufficient nitrogen plasma treatment is not favorable for the perfect growth of GaN films on the surface of Ga droplets. Therefore, the 4 nm GaN films obtained after 10 minutes of treatment is employed in subsequent microstructure observation, optical analysis, and electronic devices fabrication, etc.

The uniform thickness in broad regions and numerous samples supports the proposed Cabrera–Mott growth mechanism, where concurrent nitrides are formed among the whole metal interface, resulting in self-limiting growth to an accurate thickness. The mentioned approach is highly reproducible for growing large-area GaN sheets because the process was reproduced more than 50 times and always yielded identical, continuous, laterally extensive thin GaN films with reproducible features. The GaN film was initially printed on the SiO$_2$/Si substrate, but further tests demonstrated that the formation of uniform centimeter-scale semiconductor films on various substrates could be reproduced through the printing technology (Supplementary Fig. 3), indicating that the presented construction approach is suitable to deposit GaN on several materials. Moreover, it should be pointed out that this method is also applicable to the fabrication of GaN heterostructures. In order to be compatible with silicon-based electronic technology, the ultrathin quasi-2D film printed on SiO$_2$/Si surface should be employed for further description and device construction.

It should be noted that, Supplementary Fig. 4 presents the comparison of the temperatures and the thicknesses of GaN of our present printing technique with other classical techniques in the field of the preparation of ultrathin GaN in previous studies. Various processes have been introduced to the growth of GaN, such as metal-organic chemical vapor deposition (MOCVD), metal-organic vapour-phase epitaxy (MOVPE), hydride vapor phase epitaxy (HVPE), atomic layer deposition (ALD), plasma-enhanced atomic layer deposition (PEALD), microwave plasma-assisted atomic layer deposition (MPALD), plasma-assisted atomic layer deposition (PA-ALD), plasma-assisted molecular beam epitaxy (PAMBE), laser molecular beam epitaxy (LMBE), flow modulation epitaxy (FME), pulsed direct current (DC) sputtering and RF magnetron sputtered etc. The above processes basically require high temperature and vacuum conditions, which virtually increases the cost, and is not conducive to large-scale pervasive application, nor can it easily realize the fabrication of flexible devices, which seriously hinders the wide practices and research progress of the ultrathin GaN in flexible devices. Compared with the above technology, the room temperature GaN printing process based on liquid metal owns the advantages of simple and stable process, fast, large area, low cost, high efficiency, easy removal of excess metal, and the final sample surface is extremely clean.

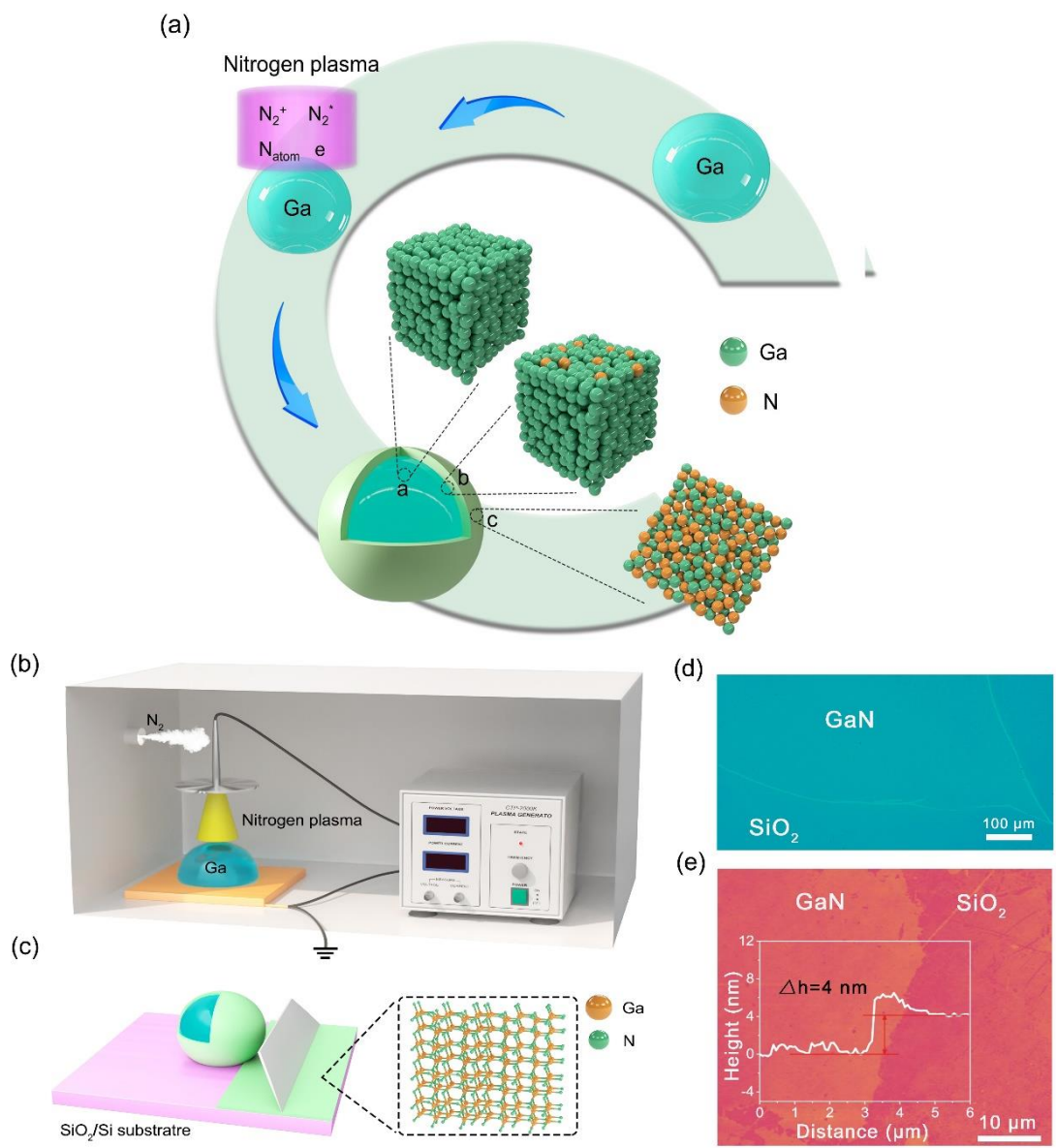

**Fig. 2 Schematic of the quasi-2D GaN synthesis and printing process.** (a) Schematic illustration of formation of surface nitride layer on liquid metals Ga via nitrogen-plasma reaction mechanism. a: Ga featuring disordered atoms, b: Ga-N bond formation at the liquid Ga surface, c: GaN layer. (b)

Setup for the self-constructed plasma system used for producing GaN from liquid metal Ga. (c) Schematic description of the vdW printing method of GaN film. The surface nitride of liquid Ga attaches to the substrate and leaves GaN on the surface. The figure on the right summarizes the crystal structure of GaN film. (d) An optical image of the printed GaN layer with some square millimeters in lateral size. (e) AFM image of a printed GaN layer on a $SiO_2$/Si wafer. The inset presents a step height profile with a 4 nm thickness.

**Properties of the ultrathin quasi-2D GaN.** Transmission electron microscopy (TEM) can be utilized to characterize the crystallographic features of the printed GaN. The GaN films were immediately moved to a TEM grid since it was printed. A high-resolution TEM micrograph (HRTEM) and selective area electron diffraction (SAED) pattern of the GaN are presented in Fig. 3a and 3b, respectively, which confirms the crystallization of the printed GaN in the polymorph. The atomic spacing of 0.285 nm in the HRTEM image and the distance of the fringes at 0.248 nm correspond to the (100) and (101) planes of GaN. The collected lattice parameters show that the GaN sample are wurtzite structural.

The GaN film's phonon modes also change compared to those of the bulk[19]. Based on the Raman spectra presented in Fig. 3c, two peaks at 561.77 and 729.59 cm$^{-1}$ appear in printed GaN, respectively, corresponding to characteristic $E_2$ (high) located at 564.30 cm$^{-1}$ and $A_1$ (LO) located at 730.34 cm$^{-1}$ modes in the bulk phase. Moreover, the above distinction reflects the change in GaN's phonons modes in quasi-2D limit. Obviously, the $E_2$ (high) and $A_1$ (LO) peaks of GaN film all show blue-shifted comparing to those of the bulk polycrystalline GaN. Generally speaking, the parameters influencing the Raman scattering include material size, order, internal stress and structural defects. The tensile (compressive) strain in GaN films will result in a blue (red) shift in the Raman spectroscopy of the GaN film. Chen et al. reported the $E_2$ peak of 2D GaN is obviously blue-shifted compared with that of the bulk, due to the tensile-strain state in the 2D limit[15]. In the present work, the thickness of our printed GaN is 4 nm, which is close to the thickness of the reported 2D GaN[15]. The tensile- strain state may also exist in the prepared GaN film, as a consequence, the $E_2$ peak and $A_1$ peak all appear blue shift for the quasi-2D GaN film. The printed GaN film for Raman spectra test is on $SiO_2$/Si substrate, it is no doubt that the Raman spectra of GaN is influenced by that of the $SiO_2$/Si substrate. As a consequence, a large bulge exists in the Raman spectrum of the 2D GaN, while it is absent in the bulk GaN.

X-ray photoelectron spectroscopy (XPS) is utilized to attain the printed GaN's chemical bonding states. Figures 3d and e present the spectra of Ga 2p and N 1s areas for the GaN, respectively. The doublet in the Ga 2p region corresponds to the $2p_{3/2}$ and $2p_{1/2}$ orbital of Ga, the characteristic gallium peak for $Ga_2O_3$ placed at ~20.4 eV was not seen, indicating Ga's quantitative transformation. The principal broad N 1s peak centered at ~397.6 eV corresponds to the N 1s region compatible with the desired N 1s area presented in GaN. Energy-dispersive X-ray spectroscopy (EDS) mapping study of the resulting film composition was shown in Supplementary Fig. 5. The GaN films for scanning electron microscope (SEM) and EDS test was scraped from $SiO_2$/Si substrate and transferred to the surface of the transparent PET tap, in order to avoid effect of abound Si on the results of GaN film's EDS. The detected element C and O in EDS results come from PET tap, which is content of element C, O and H. During the process of scraping GaN films, the uniformity of GaN film is destroyed and some films were wrinkled, so the distribution of N and Ga in observed GaN films is uneven. But the atomic ratio of Ga:N is approximate to 1:1, proving that the obtained material is GaN. The attained XPS data and the other characterization techniques aforementioned above provide the outcome that all of our printed GaN have good consistency and can be used for further empirical electronic device fabrications.

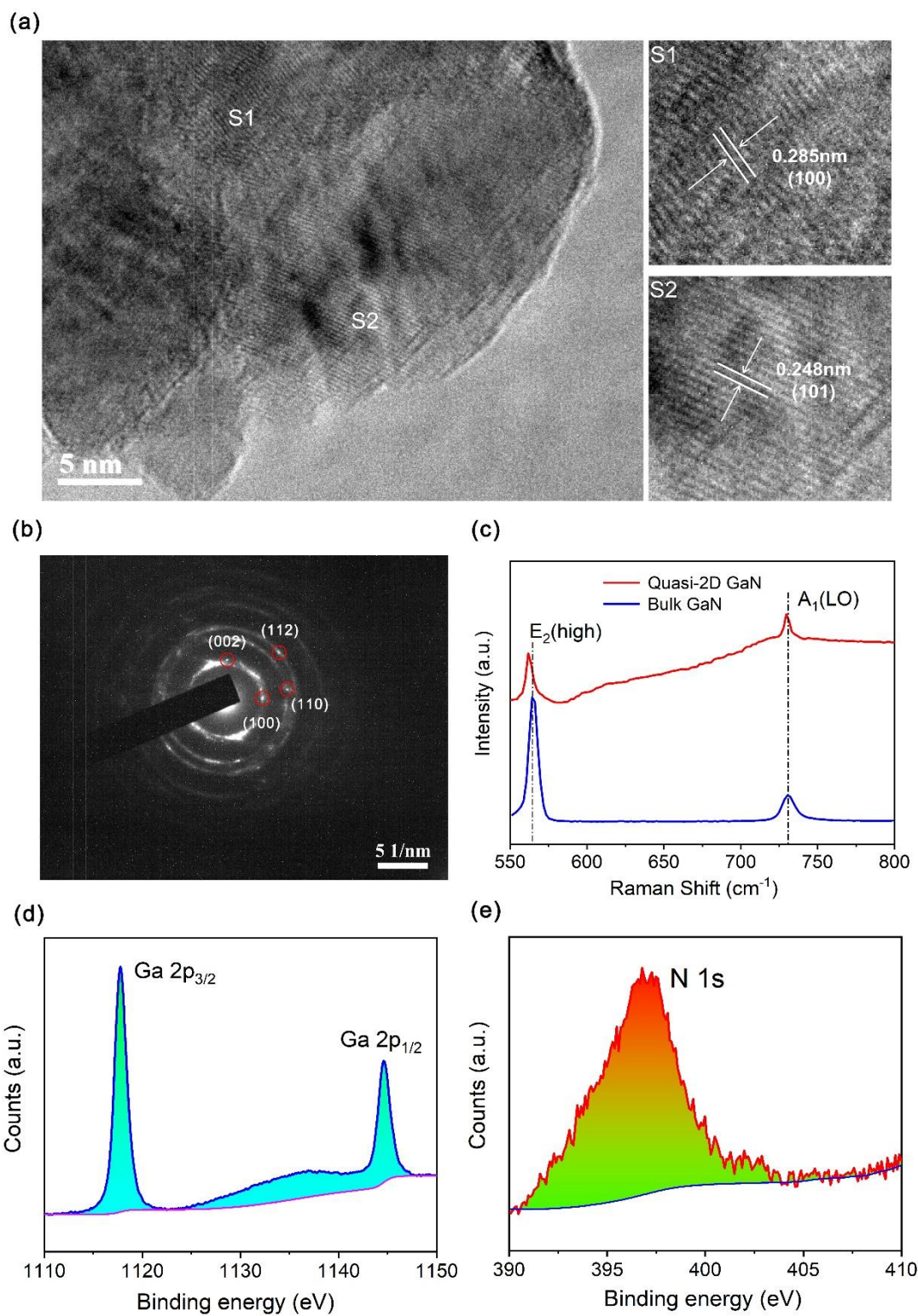

**Fig. 3 Description of the printed ultrathin quasi-2D GaN.** (a) High-resolution TEM images of GaN directly laminated to a TEM grid. (b) Indexed chosen region's diffraction pattern for GaN, demonstrating the polycrystalline structure of GaN. (c) Raman spectra for the bulk GaN and the GaN with a 4.0 nm thickness, which present the areas of the $E_2$ and $A_1$ model peaks for the sample. (d, e) XPS results of the GaN for the selected areas, (d) Ga 2p and (e) N 1s. The characteristic doublets for the Ga 2p area, $2p_{1/2}$, and $2p_{3/2}$, are located at ∼1144.8 and ∼1118.1 eV. The broad N 1s peak located at ∼397.6 eV corresponds to the desired binding energy for nitrogen in GaN.

The obtained electronic band structure and density of states (DOS) of an individual unit cell of the printed ultrathin quasi-2D GaN are shown in Fig. 4a. Vienna Ab initio Simulation Package (VASP, version: 5.4.4) combined with the projector augmented wave (PAW) approach were utilized to accomplish the first-principles computations[20–22]. The Perdew-Burke-Ernzerhof (PBE) functional integrated with the DFT-D3 correction was employed to treat the exchange-functional. The plane wave's cut-off energy was adjusted at 520 eV[23]. The Brillouin zone integration was accomplished with 15*15*6 Monkhorts-Pack point sampling to optimize the bulk GaN. The self-consistent computations can give a 10-4 eV convergence energy threshold. The optimal values of the equilibrium geometries and lattice constancies were obtained with maximum stress on all atoms in 0.01 eV/Å. For the GaN (110) surface structure, we use the 7*7*1 K-points for structural optimization and self-consistent calculations. Because the PBE functional will underestimate the band gap of the semiconductor, we also use the hybrid functional method (HSE06)[24] to calculate the band gap and DOS. Our density PBE functional analysis indicates that the GaN has a 1.51 eV direct bandgap. According to the HSE06 calculation, the printed GaN has a 3.32 eV direct bandgap. From Fig. 4b, the measured band gap derived from the UV-Vis absorption is 3.3 eV, which agrees well with the value derived from HSE06 calculation. By comparison, the calculated bandgap using energy density PBE functional analysis is smaller than the measurement result. Figure 4c shows a periodic slab of GaN with a non-polar (110) surface sliced from the wurtzite bulk phase. Generally speaking, the bandgap of GaN can be modulated up to 5 eV down to monolayer limit[25, 26]. Notably, the measured band gap of the printable GaN film (4 nm) is 3.3 eV, which is smaller than 3.5 eV of the reported 2D GaN film (1.3 nm)[14]. It is because the obtained ultrathin quasi-2D GaN is slightly thicker than 2D GaN. With the thickness increasing, the band gap of the ultrathin quasi-2D GaN decreases[15], which is close to band gap of the bulk GaN.

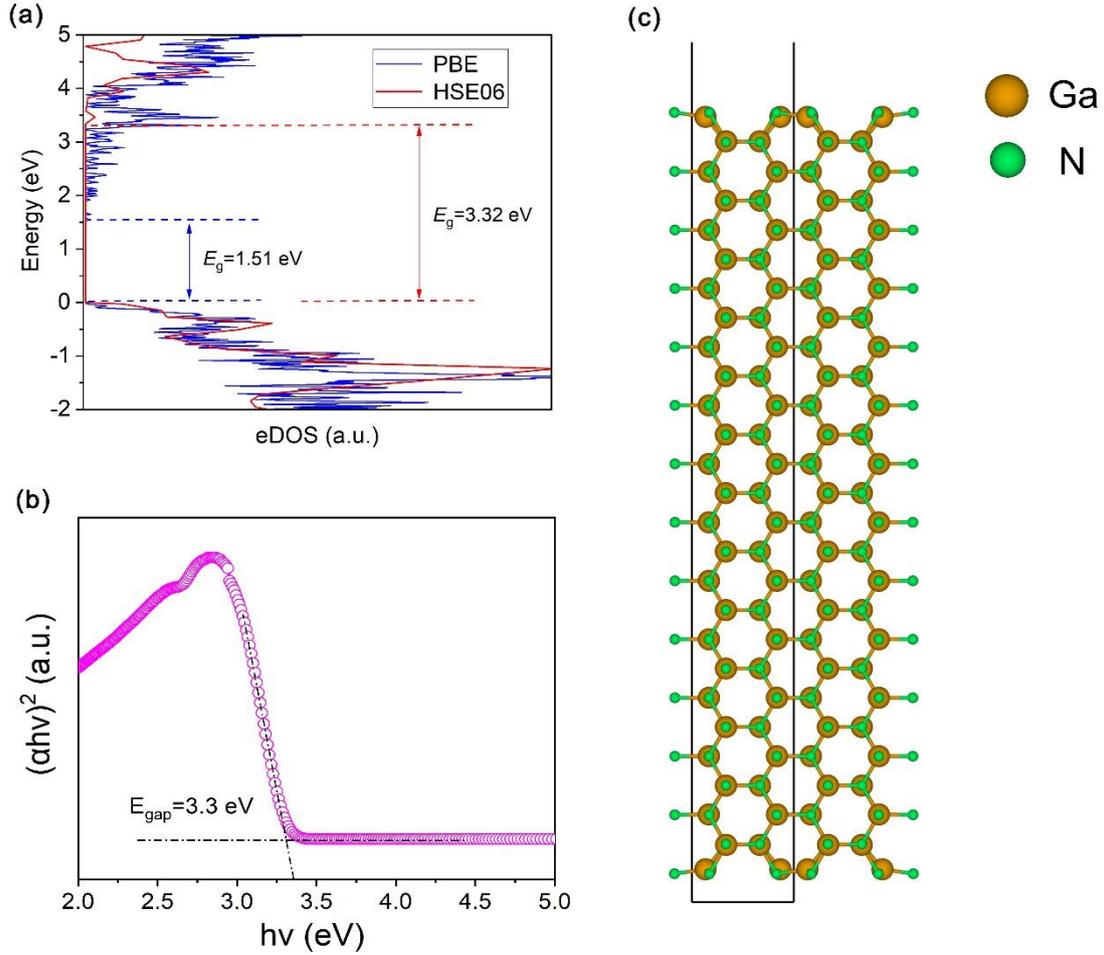

**Fig. 4 Electronic band descriptions of the printed ultrathin quasi-2D GaN.** (a) Calculated electronic DOS of printed GaN using PBE and HSE06 functional. (b) Tauc plot employed for ascertaining the electronic bandgap for GaN revealed a bandgap of 3.3 eV. (c) GaN (110) slab.

**Application of printed ultrathin quasi-2D GaN in electronic devices.** For experimental assessment of the electron transport features of the printed GaN and to investigate the potential for electronic devices, field-effect transistors (FETs) were constructed to assess the GaN for electronic device applications. Figure 5a shows the structure diagram of the transistor based on the printed GaN. We have adopted an individual side-gate design for all of the devices fabricated in this study, and detailed description of the fabrication process will be presented in the Methods section. Figure 5b presents a scanning electron microscopy (SEM) image of the device. Ag was used as gate electrodes and source–drain metal contacts, and FET channels were patterned with a width of $W_{ch}$ = 1000 μm and the length of $L_{ch}$ = 25 μm. Electrical measurements were carried out for the printed side-gated GaN FET. Figure 5c presents the transfer (drain current, $I_{ds}$, with respect to the gate voltage, $V_{gs}$) features of a representative GaN FET. Figure 5d presents the $I_{ds}$ with respect to the drain–source voltage ($V_{ds}$) at various values of $V_{gs}$ applied to the device. As presented in Fig. 5c and d, the printed GaN FET devices' p-type switching feature with an on/off ratio is more than $10^5$. The sub-threshold swing (SS) for the FET was 98 mV per decade, near the desired action. The average value of the room-temperature field-effect mobility (μ) was obtained as 53.1 cm$^2$ V$^{-1}$ s$^{-1}$, with a mobility of 57 cm$^2$ V$^{-1}$ s$^{-1}$ for the device with the best performance (The SS and mobility computations are presented in the Supplementary Note 6). A statistical analysis on the performance of many FET devices fabricated using the presently reported wafer-scale printing process was conducted. Electrical features of thirty printed GaN FETs were evaluated. The mean log ON/OFF current ratio, mobility, and SS were 5.31 ± 0.52, 53.1 ± 4.5 cm$^2$ V$^{-1}$ s$^{-1}$, and 97.6 ± 2.42 mV dec$^{-1}$ (Fig. 5e-g), respectively. A desired output was seen between various devices, considering that the mentioned devices were constructed in an academic laboratory. This uniformity yields confidence in employing the mentioned devices and methods in numerous applications like integrated circuits and active-matrix back-planes for displays. Moreover, the GaN-based FETs were significantly stable, having a high cyclability with a stable on/off ratio and stable on currents for more than 100 switching cycles under ambient conditions (Supplementary Fig. 6).

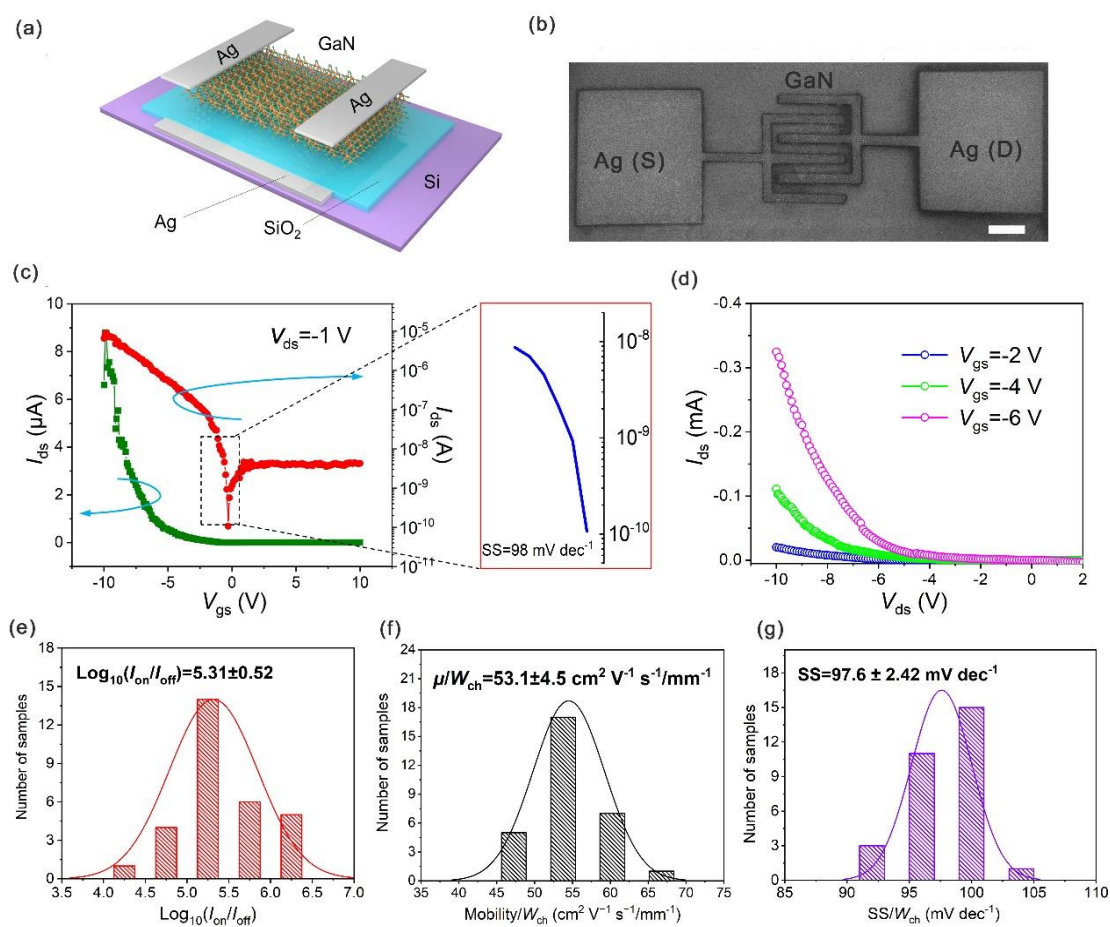

**Fig. 5 The printed GaN field-effect transistor measurements.** (a) Schematic of the device presenting a Si side-gated FET with silver source–drain electrodes. (b) The device's SEM image of the device. The scale bar corresponds to 100 μm. (c) Corresponding I$_{ds}$–V$_{gs}$ curve at V$_{ds}$ = − 1 V, showing a high I$_{on}$/I$_{off}$ ratio more than 10$^5$, small subthreshold slope of SS = 98 mV dec$^{-1}$, yielding a field-effect mobility of μ = 53.1 cm$^2$ V$^{-1}$ s$^{-1}$. (d) Set of I$_{ds}$–V$_{ds}$ output curves from an individual FET, showing a large on current density of > 0.3 mA. Histograms of the (e) Log of the ON/OFF current ratio, (f) Field-effect mobility and (g) Subthreshold slope (SS) for 30 FETs.

Notably, the observed field-effect mobility is greater compared to that of traditional high-efficiency broad-bandgap GaN-based devices[14, 27]. However, it should be noted that most of the reported GaN FET devices adopted AlGaN/GaN heterojunction as carrier transport layer, and the devices have complex structures. Because the 2D electron gas (2DEG) at the interface of AlGaN/GaN heterostructure has very high electron mobility, the performance of AlGaN/GaN heterostructure FET is better than that of the device composed of GaN film only, the two kinds of devices can not be directly compared. Therefore, for further broader comparison, we compare the performance of GaN and other p-type and n-type FETs reported and the devices in this study, Fig. 6 plots SS versus mobility for other FETs reported in the literature as well as the device from this study, the value of SS is comparable to that appeared in some of the best p-type and n-type oxide semiconductors like indium oxide, this highlights the fact that our approach offers a high current mobility while maintaining a small SS, which is mainly attributed to the high-quality electron-level GaN semiconductor. Besides, no detectable performance degradation was observed in the devices while working under ambient situations without encapsulation. This highlights that the printed GaN is of exceptional quality while indicating that additional enhancements can be achieved via the enhanced device construction. Future studies should concentrate on constructing the devices that can integrate separately addressable transistors into more complicated circuits. Potential approaches can involve using vdW heterostructures with dielectrics like hexagonal boron nitride or $Ga_2O_3$ combined with top gates[28, 29]. The mentioned device configurations can be employed to determine essential parameters like the pinch-off voltage that can be informative for incorporating future devices into functional circuits.

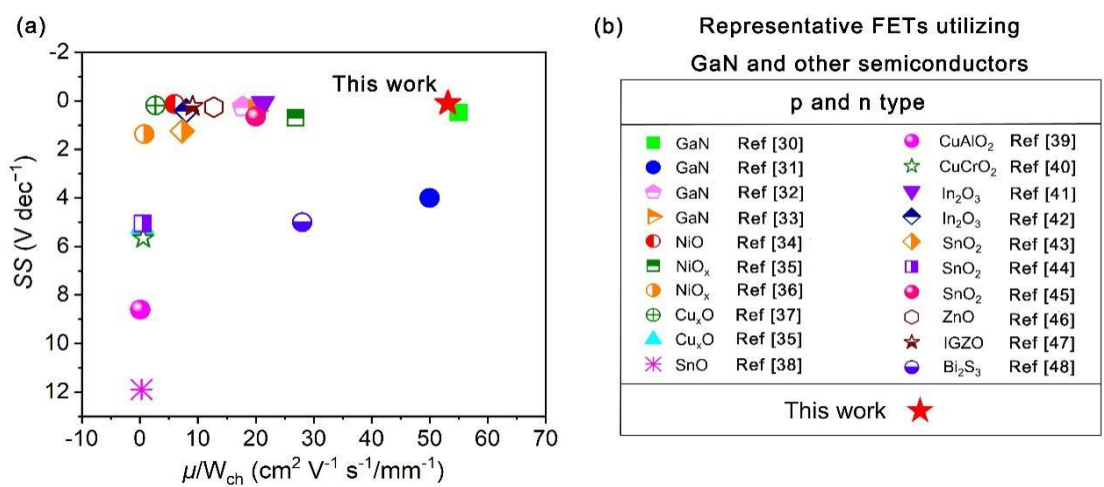

**Fig. 6 Field-effect transistor measurements of printed GaN.** (a) Plot of subthreshold slope versus field-effect mobility to compare representative FETs reported in the literature. Our device shows the highest mobility while maintaining a small SS. (b) Legend for previously reported FETs utilizing p-type or n-type semiconductors for all plots.

**3. Conclusions**

Liquid metals have rather diverse unconventional functionalities. One of the essential value of such materials is that they can serve as a reactive and templating media simultaneously and thus to synthesis and print a wide range of 2D semiconductor materials with extensive practical values. In the present work, we demonstrated the successful room temperature printing of the high-quality ultrathin quasi-2D GaN films with a large scale via a plasma nitridation reaction and through transferring occurring in the liquid metal's nitride skin constructed under this specific condition.

Moreover, complete explanation and analysis have been performed on the electrical features of the printed GaN films with thickness spanning from 1nm to several nanometers, giving essential advice for further evolution of liquid metal printed high-performance semiconductors and indicating their considerable potential for future nanoelectronics.

As example, we effectively presented the printed FET devices using GaN for the first time, depicting significant field-effect mobility and relatively small values of SS, and demonstrating an extremely steep subthreshold voltage switching behavior. This high-efficiency, simple, large-size, and cheap production process provides a new way for advancing GaN devices for power electronics applications. Moreover, the 2D or qusia-2D semiconductor can be produced as the favored material in printing different electronic devices due to the mentioned advantages. This sets up a new standard for subsequent electronics, sensors, and more practical devices. It also offers a pathway toward employing printed ultrathin 2D high-efficiency semiconductors to construct novel electronic devices using liquid metal-enabled techniques in the near future.

## 4. Methods

**Materials.** Gallium with a purity of >99.99% was bought from Sigma Aldrich and utilized without additional refinement. The remaining materials were bought from usual suppliers and employed without refinement.

**Formation of GaN on liquid gallium by nitridation reaction.** In order to obtain GaN films with high purity and low defects, the whole preparation and printing process were carried out in a glovebox in a pure $N_2$ environment at a pressure of 1 atm ~ 3 atm. The $O_2$ content in the glove box atmosphere is maintained below 3 ppm and the $H_2O$ content is controlled below 0.5 ppm. High purity Ga was melted and then placed in NaOH solution to remove the oxide scale on the surface. The clean liquid Ga was extracted to 10 ml by a syringe and place it on the surface of the low-voltage electrode stainless steel plate of the plasma trigger device, and the stainless steel plate is grounded. In the environment of high purity $N_2$, it can be seen that the surface of liquid Ga is bright without any oxide film. The stainless steel disk wrapped in transparent quartz of plasma high-voltage electrode is suspended vertically above liquid Ga, and the distance between the disk and the surface of liquid Ga is 1mm. A voltage of 50 kV is applied between the two electrodes of the plasma trigger device through the voltage regulating device. At this time, the electric field strength of $N_2$ breakdown is $5 \times 10^7$ V m$^{-1}$, the output current is 16 A, and the uniform and dense GaN skin can be obtained on the liquid Ga surface after glow discharge treatment for a certain time.

**Printing process of ultrathin quasi-2D GaN films.** Wafers of 500 nm $SiO_2$ on Si ($SiO_2$/Si) were washed by the deionized (DI) water for 1 min, and then sonicated in acetone and isopropyl alcohol (IPA) for 10 min (25 °C) and blown dry with $N_2$ gas. The 3 min of oxygen ($O_2$) plasma (Emitech K-1050X) at 100 W was then applied to wafers under the low vacuum (0.6 Torr). The nitrided Ga droplets was placed on the $SiO_2$/Si substrate and execute the complete printing program. The size of nitride film changes with the droplet diameter and the travel distance of the scraper. The nitride layer formed on the liquid Ga's surface can be transformed and printed on the substrate through gently scraping the droplets from one end of the substrate to the other end with a scraper. An extra force within the extrusion phase can damage the nitride layer. By this extrusion printing method, high-quality GaN films with a transverse size greater than a few centimeters can be effectively printed on the substrate.

**Mechanical and chemical cleaning process.** In order to remove all liquid metal parts that remained on the sample, a facile mechanical ethanol cleaning method was employed. At first, about 100 ml of ethanol were taken in a beaker, and the beaker was heated on a heating plate to 100 °C. Then, the substrate with printed nitride film was immersed in hot ethanol with tweezers. In order to eliminate the metal residues, a swab tool was utilized for wiping the substrate immersed in ethanol. Due to a strong vdW adhesion between the nitride film and the bottom layer, the nitride film still sticks to the silicon oxide surface within the wiping proceeding. Besides, the weak adherence between the deposited nitride layer, the liquid metal, and the film could be quickly removed to maintain the film clean and intact. Moreover, a chemical process was employed to clean the samples for a complete elimination of the metal residue on the substrate. An Iodine/triiodide ($I^-/I^{3-}$) solution (100 mmol $L^{-1}$ LiI and 5 mmol $L^{-1}$ $I_2$) was constructed in ethanol and then located on a hot plate to heat up to 50 °C. In order to eliminate metallic inclusions, a substrate printed with the GaN film was immersed in a heated $I^-/I^{3-}$ solution for a time interval. At last, the residual etchant was removed by cleaning the sample in deionized water. The liquid metal particles can be successfully eliminated using the mentioned two cleaning processes.

**FET fabrication.** In order to construct fully printed side-gated GaN FETs, firstly, part of the GaN/SiO$_2$ region is etched with HF solution with concentration of 1mol $L^{-1}$ for about 10 s. Then, the etching area was cleaned with ethanol, the clean Si of a certain area was obtained on the substrate. The Ag ink (Ag40X, UT Dots, Inc.) involved 40 wt % Ag nanoparticles, with about 20 nm particle diameters, dispersed in a solvent mixture of xylene and terpineol (9:1 by volume). The constructed ink was printed on the GaN films and SiO$_2$/Si substrates. A scientific 3B inkjet printer from Prtronic was adopted to verify the inkjet printing details. The obtained sample was located on the inkjet printer's panel at room temperature. A target image file was applied to the computer, which could be converted into a printable file through the software. Under computer control, source/drain electrodes and side-gate electrodes were subsequently patterned by printing Ag on the substrate. Finally, the printed samples were then sedimented at 120 °C in the air for 30 min in a furnace (MDL 281, Fisher Scientific Co.) to improve the conductivity.

**Characterization.** The AFM images were employed by a Bruker Dimension Icon with "Scanasyst-air" AFM tips. A JEOL 2100F TEM/STEM (2011) system working at a 200 kV acceleration voltage involving a bright-field Gatan OneView 4k charge-coupled device (CCD) camera was utilized for both the low-resolution HRTEM imaging and SAED. A laser micro-Raman spectrometer (Renishaw in Via, 532 nm excitation wavelength) was adopted to accomplish the Raman spectroscopy. Moreover, the energy dispersive X-ray spectroscopic (EDS) measurements were used to collect the elemental mapping of the as-prepared samples. A thermo Scientific K-alpha XPS spectrometer associated with monochromatic X-rays from an Al anode (hv=~1486.6 eV) was adopted to perform the XPS analysis. A UV–visible (UV–vis) absorbance spectrometer (Hitachi U3900 UV–vis spectrophotometer) was utilized to evaluate the film's optical bandgaps. A Cascade Microtech Summit 12000 semiautomated probe station linked to a Keithley 4200 Semiconductor Device Analyzer was adopted to measure GaN FETs at room temperature.

**Data availability**

The data used to determine the data points shown within the plots presented in this paper, and other findings from this study, are available from the corresponding author upon reasonable request.

**Acknowledgements**

This work was partially supported by the National Natural Science Foundation of China under Grants No. 51890893 and 91748206.

**Author contributions**

The project was designed and directed by J. L. who also wrote part of the manuscript. Q. L. and B. D. D. developed the synthesis and printing procedure for GaN while also conducting the structure and properties measurements. B. D. D., B. Y. X., D. K. W. and J. F. Y. designed nitrogen plasma and synthesized GaN on liquid metal Ga. Q. L. led the device fabrication with contributions from B. D. D. and J. Y. G., Q. L. and B. D. D. characterized the FET devices. Q. L analyzed the material and device characteristics and drafted the manuscript. All authors revised the manuscript.

**Competing interests**

The authors declare no competing interests.

**References**

[1] Sanders, N., Bayerl, D., Shi, G., Mengle, K. A. & Kioupakis, E. Electronic and optical properties of two-dimensional GaN from first-principles. *Nano Lett*. **17,** 7345–7349 (2017).

[2] Mansurov, V., Malina, T., Galitsyn, Yu. & Zhuravlev, K. Graphene-like AlN layer formation on (111) Si surface by ammonia molecular beam epitaxy. *J. Cryst. Growth*. **428,** 93–97 (2015).

[3] Xu, L. et al. Rationally Designed 2D/2D SiC/g-$C_3N_4$ photocatalysts for hydrogen production. *Catal. Sci. Technol*. **9,** 3896–3906 (2019).

[4] Li, Q., Xu, L., Luo, K. W., Wang, L. L. & Li, X. F. SiC/$MoS_2$ layered heterostructures: Promising photocatalysts revealed by a first-principles study. *Mat. Chem. Phys*. **216,** 64–71 (2018).

[5] Khanna, S. M., Webb, J., Tang, H., Houdayer, A. J. & Carlone, C. 2 MeV proton radiation damage studies of gallium nitride films through low temperature photoluminescence spectroscopy measurements. *IEEE Trans. Nucl. Sci*. **47,** 2322–2328 (2000).

[6] Luo, K. W., Xu, L., Wang, L. L., Li, Q. & Wang, Z. Ferromagnetism in zigzag GaN nanoribbons with tunable half-metallic gap. *Comput. Mater. Sci*. **117,** 300–305, (2016).

[7] Zhu, Y., Li, H., Chen, T., Liu, D. & Zhou, Q. Investigation of the electronic and magnetic properties of low-dimensional $FeCl_2$ derivatives by first-principles calculations. *Vacuum* **182,** 109694 (2020).

[8] Kolobov, A. V., Fons, P., Tominaga, J., Hyot, B. & André, B. Instability and spontaneous reconstruction of few-monolayer thick GaN graphitic structures. *Nano Lett*. **16,** 4849–4856 (2016).

[9] Nakamura, S. Nobel Lecture: Background story of the invention of efficient blue InGaN light emitting diodes. *Rev. Mod. Phys*. **87,** 1139 (2015).

[10] Xiao, W. -Z. et al. Ferromagnetic and metallic properties of the semihydrogenated GaN sheet. *Phys. Status Solidi B* **248,** 1442–1445 (2011).

[11] Zhang, J., Sun, C. & Xu, K. Modulation of the electronic and magnetic properties of a GaN nanoribbon from dangling bonds. *Sci. China Phys. Mech. Astron*. **55,** 631–638 (2012).

[12] Kandalam, A. K. et al. First principles study of polyatomic clusters of AlN, GaN, and InN. 1. Structure, stability, vibrations, and ionization. *J. Phys. Chem. B* **104,** 4361–4367 (2020).

[13] Onen, A., Kecik, D., Durgun, E. & Ciraci, S. Onset of vertical bonds in new GaN multilayers:

beyond van der Waals solids. *Nanoscale* **10,** 21842–21850 (2018).

[14] Syed, N. et al. Wafer-sized ultrathin gallium and indium nitride nanosheets through the ammonolysis of liquid metal derived oxides. *J. Am. Chem. Soc.* **141,** 104–108 (2019).

[15] Chen, Y. et al. Growth of 2D GaN single crystals on liquid metals. *J. Am. Chem. Soc.* **140,** 16392−16395 (2018).

[16] Balushi, Z. Y. Al. et al. Two-dimensional gallium nitride realized via graphene encapsulation. *Nat Mater.* **15,** 1166–1171 (2016).

[17] Sakakura, T., Murakami, N., Takatsuji, Y., Morimoto, M. & Haruyama, T. Contribution of discharge excited atomic N, $N_2^*$, and $N_2^+$ to a plasma/liquid interfacial reaction as suggested by quantitative analysis. *ChemPhysChem* **20,** 1467–1474 (2019).

[18] Le, K. et al. Controllably doping nitrogen into 1T/2H $MoS_2$ heterostructure nanosheets for enhanced supercapacitive and electrocatalytic performance by low-power $N_2$ plasma. *ACS Appl. Mater. Interfaces* **13**, 44427–44439 (2021).

[19] Caldwell, J. D. et al. Atomic-scale photonic hybrids for mid-infrared and terahertz nanophotonics. *Nat. Nanotechnol.* **11,** 9–15 (2016).

[20] Kresse, G. & Furthmüller, J. Efficient iterative schemes for ab initio total-energy calculations using a plane-wave basis set. *Phys. Rev. B* **54,** 11169 (1996).

[21] Paier, J., Marsman, M., Hummer, K. & Kresse, G. Screened hybrid density functionals applied to solids. *J. Chem. Phys.* **124,** 154709 (2006).

[22] Blöchl, P. E. Projector augmented-wave method. *Phys. Rev. B* **50,** 17953, (1994).

[23] Jain, A. et al. Commentary: The materials project: A materials genome approach to accelerating materials innovation. *APL Materials* **1,** 011002 (2013).

[24] Heyd, J. & Scuseria, G. E. Hybrid functionals based on a screened Coulomb potential. *J. Chem. Phys.* **118,** 8207–8215 (2003).

[25] Wang, W. et al. Lattice Structure and Bandgap Control of 2D GaN Grown on Graphene/Si Heterostructures. *Small* **15**, 1802995 (2019).

[26] Al Balushi, Z. et al. Two-dimensional gallium nitride realized via graphene encapsulation. *Nature Mater* **15**, 1166–1171 (2016).

[27] Zheng, Z. et al. Gallium nitride-based complementary logic integrated circuits. *Nat Electron* **4,** 595-603 (2021).

[28] Wurdack, M. et al. Ultrathin $Ga_2O_3$ glass: A large-scale passivation and protection material for monolayer $WS_2$. *Adv. Mater.* **33,** 2005732 (2020).

[29] Lee, G.-H. et al. Highly stable, dual-gated $MoS_2$ transistors encapsulated by hexagonal boron nitride with gate-controllable contact, resistance and threshold voltage. *ACS Nano* **9,** 7019–7026 (2015).

[30] Gupta, C. et al. Comparing electrical performance of GaN trench-gate MOSFETs with a-plane (1120) and m-plane (1100) sidewall channels. *Appl. Phys. Express* **9**, 121001 (2016).

[31] Zhang, K., Sumiya, M., Liao, M., Koide, Y. & Sang, L. P-Channel InGaN/GaN heterostructure metal-oxide semiconductor field effect transistor based on polarization-induced two-dimensional hole gas. *Sci Rep* **6**, 23683 (2016).

[32] Liu, C., Khadar, R. A. & Matioli, E. GaN-on-Si Quasi-Vertical Power MOSFETs. *IEEE Electron Device Lett* **39**, 71–74 (2018).

[33] Gupta, C. et al. OG-FET: An in-situ Oxide, GaN interlayer based vertical trench MOSFET. *IEEE Electron Device Lett* **37**, 1601–1604 (2016).

[34] Xu, W. et al. P-type transparent amorphous oxide thin-film transistors using low-temperature solution-processed nickel oxide. *J. Alloys Compd.* **806,** 40–51 (2019).

[35] Shan, F. et al. High-mobility p-type $NiO_x$ thin-film transistors processed at low temperatures with $Al_2O_3$ high-k dielectric. *J. Mater. Chem. C.* **4,** 9438–9444 (2016).

[36] Hu, H., Zhu, J., Chen, M., Guo, T. & Li, F. Inkjet-printed p-type nickel oxide thin-film transistor. *Appl. Surf. Sci.* **441,** 295–302 (2018).

[37] Liu, A. et al. In situ one-step synthesis of p-type copper oxide for low-temperature, solution processed thin-film transistors. *J. Mater. Chem. C.* **5,** 2524–2530 (2017).

[38] Yim, S. et al. Lanthanum doping enabling high drain current modulation in a p-type tin monoxide thin-film transistor. *ACS Appl. Mater. Interfaces* **11,** 47025–47036 (2019).

[39] Li, S. et al. Preparation and characterization of solution-processed nanocrystalline p-type $CuAlO_2$ thin-film transistors. *Nanoscale Res. Lett.* **13,** 259 (2018).

[40] Nie, S. et al. Solution-processed ternary p-type $CuCrO_2$ semiconductor thin films and their application in transistors. *J. Mater. Chem. C.* **6,** 1393-1398 (2018).

[41] Ding, Y. et al. High-performance indium oxide thin-film transistors with aluminum oxide passivation. *IEEE Electron Device Lett.* **40,** 1949-1952 (2019).

[42] Leppäniemi, J., Huttunen, O. -H., Majumdar, H. & Alastalo, A. Flexography-printed $In_2O_3$ semiconductor layers for high-mobility thin-film transistors on flexible plastic substrate. *Adv. Mater.* **27,** 7168-7175 (2015).

[43] Lee, W. Y. et al. Densification control as a method of improving the ambient stability of solgel-processed $SnO_2$ thin-film transistors. *IEEE Electron Device Lett.* **40,** 905-908 (2019).

[44] Zhang, L. et al. Structural, chemical, optical, and electrical evolution of solution-processed $SnO_2$ films and their applications in thin-film transistors. *J. Phys. D: Appl. Phys.* **53,** 175106 (2020).

[45] Liang, D. -D., Zhang, Y. -Q., Cho, H. J. & Ohta, H. Electric field thermopower modulation analyses of the operation mechanism of transparent amorphous $SnO_2$ thin-film transistor. *Appl. Phys. Lett.* **116,** 143503 (2020).

[46] Saha, J. K. et al. Highly stable, nanocrystalline, ZnO thin-film transistor by spray pyrolysis using high-k dielectric. *IEEE Trans. Electron Devices* **67,** 1021-1026 (2020).

[47] Xu, W. et al. Low temperature solution-processed IGZO thin-film transistors. *Appl. Surf. Sci.* **455,** 554-560 (2018).

[48] Messalea, K. A. et al. Two-step synthesis of large-area 2D $Bi_2S_3$ nanosheets featuring high in-plane anisotropy. *Adv. Mater. Interfaces* **7,** 2001131 (2020).

# Supplementary information

**Supplementary Note 1: Schematic of GaN synthesis process and changes of fresh Ga droplet before and after subject to nitrogen plasma treatment**

As is well known, liquid Ga would be rapidly oxidized when exposed to ambient air, due to its high reactivity with oxygen. During the process of nitriding treatment, the self-confined gallium oxide on Ga surface will inhabit nitrogen plasma from directly hitting and reacting with Ga, resulting in the inability to generate GaN. Therefore, in order to guarantee the GaN film growth on the surface of Ga droplets, the oxide films need to be removed by using NaOH. In this way, there is no gallium oxide formation on the surface of fresh Ga droplet in pure $N_2$ atmosphere (Supplementary Fig. 1(b)), when it is placed on the low-voltage electrode (cathode) stainless steel plate of the plasma trigger device. After subject to 10 min nitrogen plasma treatment, a dense defined GaN film can be formed on the surface of Ga droplets in pure $N_2$ atmosphere (Supplementary Fig. 1(c)).

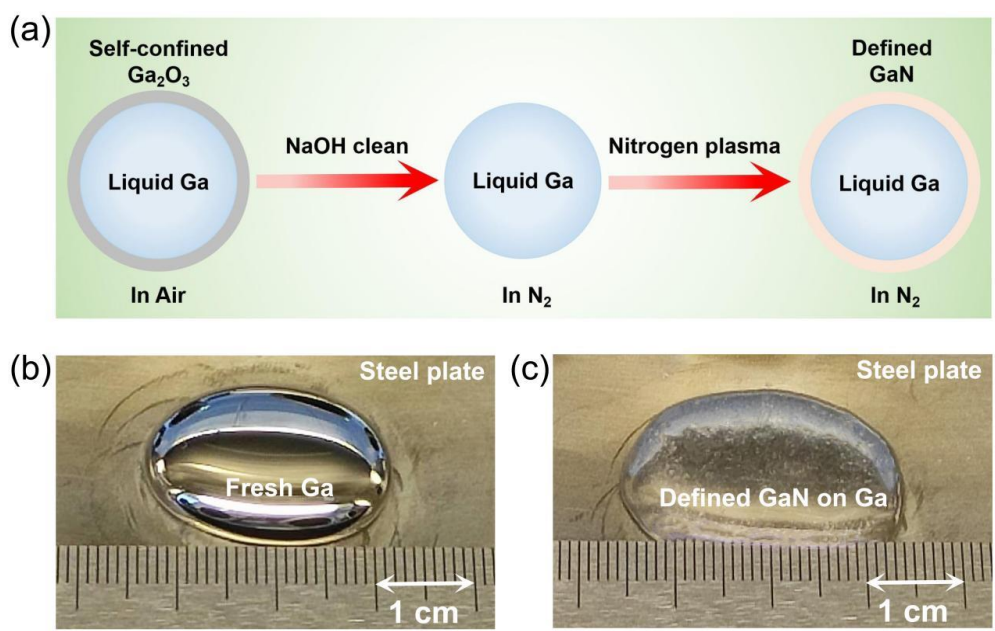

**Supplementary Figure 1.** Gas mediated fabrication of gallium base functional film. (a) Flowchart of defined GaN formation on liquid Gallium's surface. (b) Fresh liquid Gallium droplets cleaned by NaOH in $N_2$ atmosphere. (b) Gallium droplets with surface defined GaN after nitrogen plasma treatment in $N_2$ atmosphere.

**Supplementary Note 2: Printing output parameters in diverse thickness range**

    The present method has wide adaptability. Additional experiments were performed to clarify the effects of nitrogen plasma triggering time on the film thickness of 2D GaN semiconductor. Extended data from Supplementary Fig. 2 and Figure 1b shows that extending the treatment time before printing leads to no unlimited increase in sheet thickness of the 2D GaN. A higher surface coverage was obtained when a step by step coating procedure was adopted. Here the droplet allowed to nitride for a set time, is moved to the distance of one droplet diameter and then allowed to re-nitride for the same period of time prior to being moved again. For this stepped roll-printing, the nitride time per step was found to have a certain effect on substrate coverage and lateral dimensions of the isolated sheets. When the droplets were rolled at relatively slower step-speeds, a substrate coverage exceeding 50% can be obtained. This raised interesting issues for the liquid metal gallium surface confined nitridation reaction mechanisms which had never been tackled before and tremendous chances are opened to pursue in the coming time.

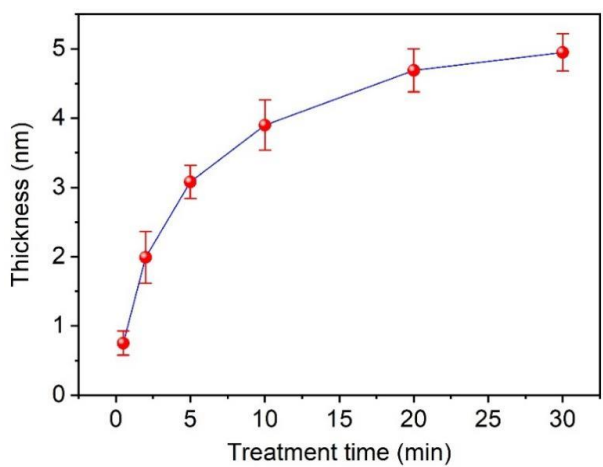

**Supplementary Figure 2.** Synthesis technique characterization on the fabricated GaN film. The thickness of nitriding layer on the surface of liquid Ga against time shows no significant change of the sheet thickness when prolonging the nitriding time.

**Supplementary Note 3: GaN with different substrates and printing layers**

As shown in Supplementary Fig. 3(a), the GaN film with thickness lower than 1 nm can be harvested via solely printing nitrided Ga droplets, when the nitrogen plasma triggering time is set as 0.5 min. The obtained GaN film can be defined as 2D GaN. However, the uniformity and continuity of prepared 2D GaN are not always good enough, due to the existing of cracks and holes involved. From Supplementary Fig. 3(a), (b) and (c), we can see dense and uniform GaN films on various substrates via 4 nm thick GaN film being repeatedly printed five times. The thickness of these films is about 20 nm, indicating the proposed printing process can easily control the thickness of synthesized GaN film and is very practical for engineering purpose.

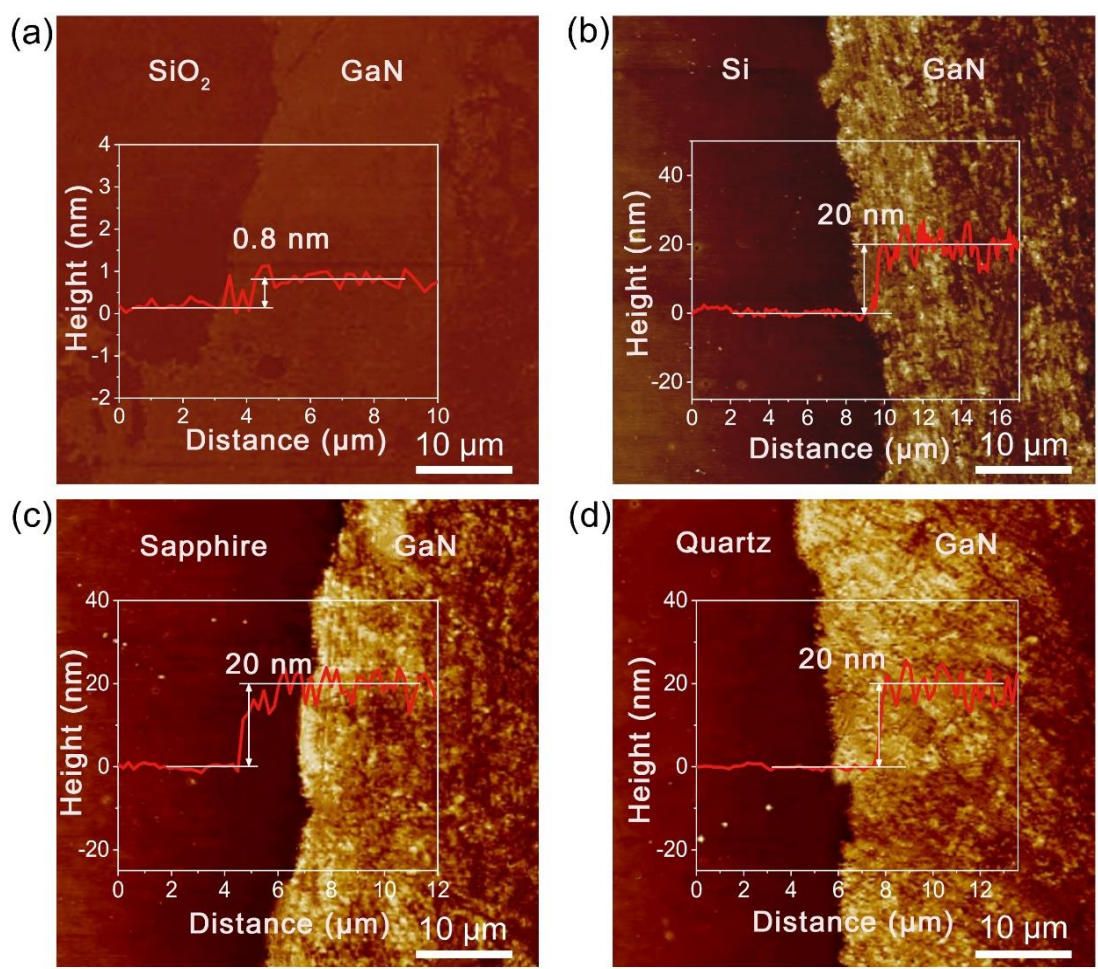

**Supplementary Figure 3.** Characterization of the printed GaN film. (a) AFM image of one touch print of the GaN film on a $SiO_2$/Si wafer. (b), (c) and (d) AFM images of frequently printed GaN films on Si wafer, sapphire and quartz, respectively.

**Supplementary Note 4: Comparison of the present printing process and other representative reported preparation process of GaN**

At present, all preparation processes of GaN basically request high temperature and vacuum conditions, which meanwhile also increases the cost, is not conducive to large-scale industrial applications, and can not easily realize flexible devices, This may seriously hinder the industrialization and research progress of ultra-thin GaN in flexible devices. Compared with the existing technologies, the room temperature GaN printing process enabled from liquid metal displayed the advantages of simple and stable process, low cost and high efficiency (Supplementary Fig. 4). Due to the low process temperature (room temperature), it can be applied to the development of various GaN based flexible electronic and optoelectronic devices in the near future.

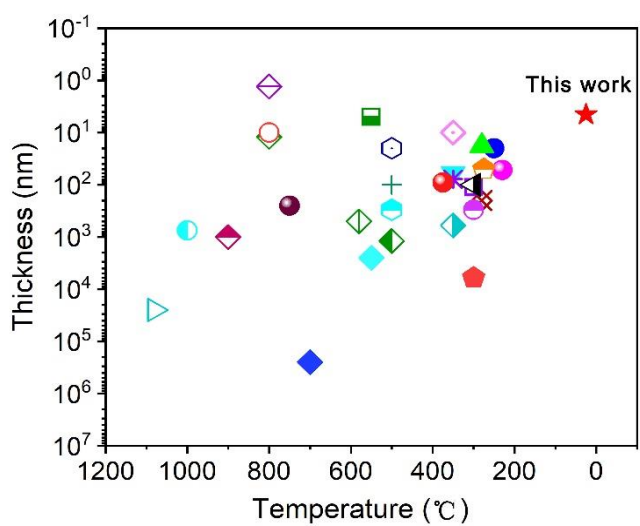

- ◆ MOCVD (*Ref.1 Opt. Express 29, 36559 (2021)*)
- ＋ MOCVD (*Ref.2 Results Phys. 24,104209 (2021)*)
- ◆ MOCVD (*Ref.3 Solid State Electron. 179,107980 (2021)*)
- ⬠ MOCVD (*Ref.4 J. Mater. Chem. C, 9, 2243 (2021)*)
- ◆ MOCVD (*Ref.5 Microsc Microanal. 25,1383 (2019)*)
- ◐ MOCVD (*Ref.6 Phys. Status Solidi A 216,1900026 (2019)*)
- ▷ MOCVD (*Ref.7 J. Am. Chem. Soc.140,16392 (2018)*)
- ⬦ MOCVD (*Ref.8 J. Am. Chem. Soc.141,104 (2019)*)
- ⬡ MOVPE (*Ref.9 J.Cryst.Growth 524,125167 (2019)*)
- ⬟ HVPE (*Ref.10 J. Electron. Mater. 50, 6688 (2021)*)
- ✳ ALD (*Ref.11 Adv. Funct. Mater.31, 2101441 (2021)*)
- ▼ ALD (*Ref.12 Chem. Mater. 33, 3266 (2021)*)
- ◆ ALD (*Ref.13 Nanomaterials 10, 2434 (2020)*)
- ● ALD (*Ref.14 J. Phys. Chem. C 123, 23214 (2019)*)
- ◆ PEALD (*Ref.15 Appl. Phys. Lett.116, 211601 (2020)*)
- ◇ PEALD (*Ref.16 Cryst. Growth Des. 21, 1778 (2021)*)
- ● PEALD (*Ref.17 Ceram. Int. 46, 5765 (2020)*)
- ✳ PEALD (*Ref.18 Acta Metall Sin-ENGL, 32, 1530 (2019)*)
- ▲ PEALD (*Ref.19 J. Mater. Chem. A, 7, 25347 (2019)*)
- ● MPALD (*Ref.20 Opt. Mater. Express 9, 4187 (2019)*)
- ⬠ PAALD (*Ref.21 J Vac Sci Technol A 37, 050901 (2019)*)
- ◇ PAMBE (*Ref.22 Appl. Phys. Lett. 117, 254104 (2020)*)
- ⬡ LMBE (*Ref.23 Phys. Scr. 96, 085801 (2021)*)
- ● MBE (*Ref.24 Vacuum 164, 72 (2019)*)
- ■ FME (*Ref.25 Semicond. Sci. Technol. 35, 095014 (2020)*)
- ● DC sputtering (*Ref.26 Coatings 9, 419 (2019)*)
- □ RF magnetron sputtering (*Ref.27 Bull. Mater. Sci. 42,196 (2019)*)
- ○ RF magnetron sputtering (*Ref.28 Emerg. Mater. Res. 8,1 (2019)*)
- ◁ RF magnetron sputtering (*Ref.29 Opt. Quant. Electron.51, 81 (2019)*)
- ★ This work

**Supplementary Figure 4.** Graphic comparison of present printing processes and GaN preparation technology in literature. New technology shows evident advantages.

**Supplementary Note 5: Distribution of elements across the GaN film**

    The chemical composition of the film surface is analyzed by EDS energy spectrum, as shown in the Supplementary Fig. 5, which shows that four elements are detected. The detected element C and O in EDS results come from PET tap, which indicates the content of element C, O and H, respectively. The existence of Ga and N elements in the spectrum proves that the components of the printed film are Ga and N atoms. The atomic percentages of the two elements are 16.24% and 18.37%, which is approximately the stoichiometric ratio of 1:1. It can thus be concluded that the sample prepared by this printing technology is pure GaN.

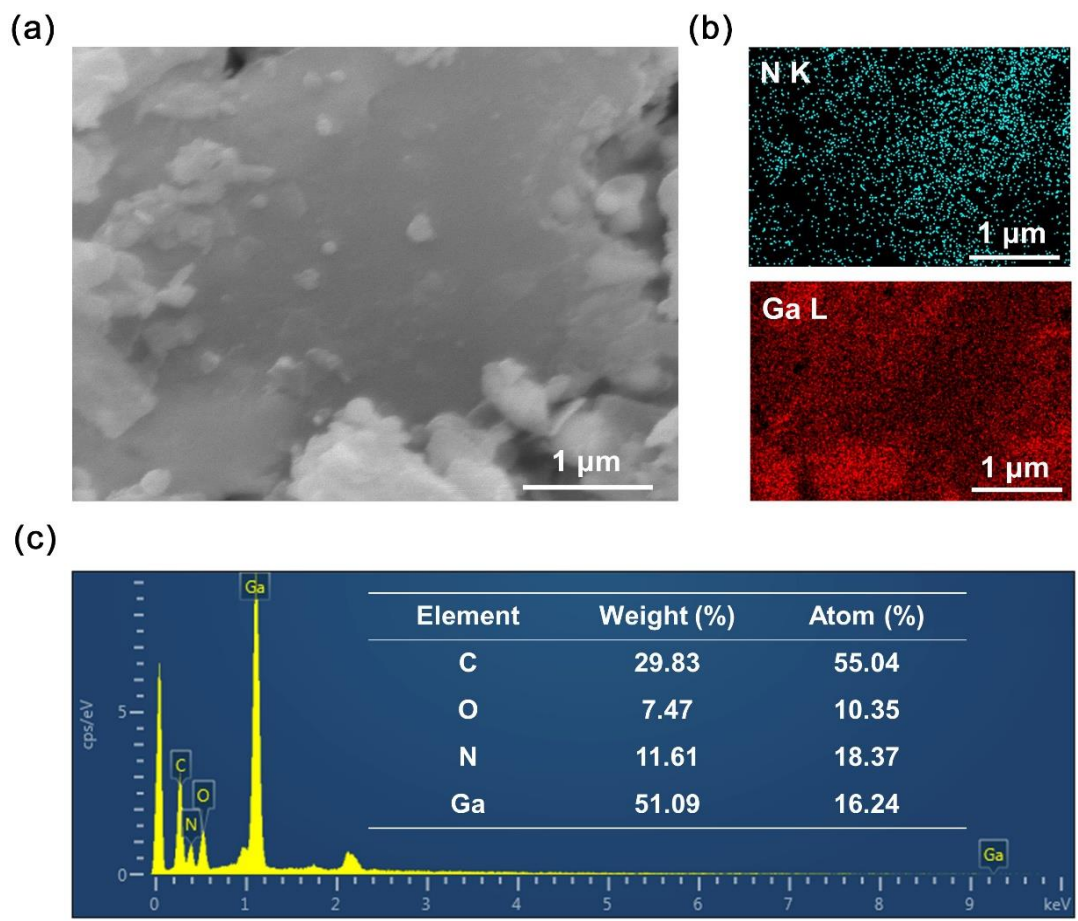

**Supplementary Figure 5.** Output of the printed GaN film. (a) SEM image of an ultrathin quasi-2D GaN sample. (b) The EDS mapping of Ga and N elements along to the GaN film harvested from the surface of nitrogen plasma triggered liquid Ga. (c) EDS spectra of the GaN by printing.

**Supplementary Note 6: Field-effect mobility and sub-threshold swing (SS) calculation**

In order to collect the carrier mobility for estimating the performance of printed FET and making a comparison with the reported GaN FETs, the field-effect mobilities were calculated using field-effect method, which is commonly used in former FETs.

In the present work, the field-effect mobilities of printed GaN-FET devices at ambient temperatures were calculated using this method, i.e.

$$\mu_{device} = \frac{L}{V_{ds} C_{ox} W} \frac{dI_{ds}}{dV_{gs}} \tag{S1}$$

Since the dielectric is 500 nm SiO$_2$, the capacitance per unit area is calculated to be:

$$C_{ox} = \frac{\varepsilon_0 \varepsilon_{ox}}{t_{ox}} = \frac{3.9 \times 8.85 \times 10^{-14} F/cm}{500 \times 10^{-7} cm} = 6.90 \times 10^{-9} F/cm^2 \tag{S2}$$

where $\mu_{device}$ is the hole mobility, $I_{ds}$ and $V_{ds}$ refer to the drain-source current and voltage, respectively, $V_{gs}$ is the gate voltage, L and W are the length and width of the channel, respectively.

When the gate bias voltage is lower than the threshold voltage, there will be leakage current flow on the semiconductor surface. Generally, this working region is called sub threshold region. In the subthreshold region, the drain current comes from the diffusion of carriers from source to drain rather than drift. The slope of the drain current curve depends on the capacitance of the gate insulator and the interface trap charge density. The reciprocal of the logarithmic slope of the source drain current to the gate voltage is called the subthreshold swing (SS), as follows:

$$SS = \frac{\partial V_G}{\partial (\log I_{ds})} \tag{S3}$$

**Supplementary Note 7: Stability of GaN-FET device.**

It is found that ultrathin quasi-2D GaN based FET is very stable, owns good recyclability, has consistent on / off ratio and stable on / off current in more than 100 switching cycles under environmental conditions (Supplementary Fig. 6). This highlights the excellent quality of the currently printed GaN, and also shows that further improvement can be achieved by improving the device design.

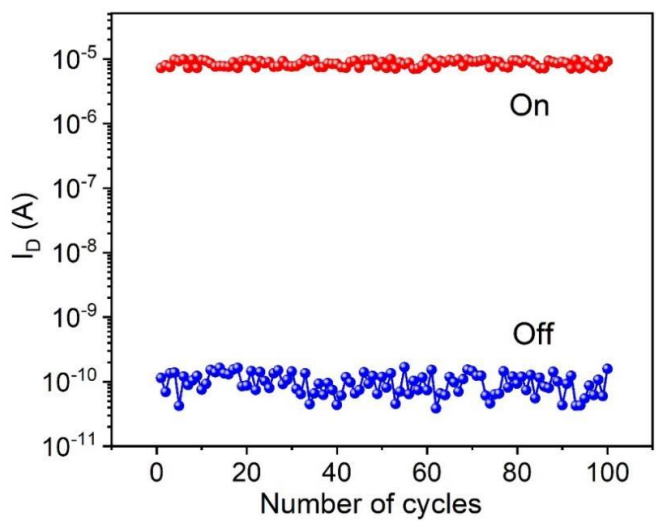

**Supplementary Figure 6.** Stability testing of a GaN field-effect transistor showing 100 On-Off cycles. No degradation in performance is observed.

**Supplementary References**

[1] Chen, D. et al. Improved electro-optical and photoelectric performance of GaN-based micro-LEDs with an atomic layer deposited AlN passivation layer. *Opt. Express* **29**, 36559-36566 (2021).

[2] Zhang, W. et al. Investigation of normally-off GaN-based p-channel and n-channel heterojunction field-effect transistors for monolithic integration. *Results Phys.* **24**, 104209 (2021).

[3] Su, C. W., Wang, T. W., Wu, M. C., Ko, C. J. & Huang, J. B. Fabrication and characterization of GaN HEMTs grown on SiC substrates with different orientations. *Solid State Electron.* **179**, 107980 (2021).

[4] Shervin, S. et al. Flexible single-crystalline GaN substrate by direct deposition of III-N thin films on polycrystalline metal tape. *J. Mater. Chem. C* **9**, 2243-2251 (2021).

[5] Saha, S. et al. Microstructural Characterization of GaN Grown on SiC. *Microsc Microanal.* **25**, 1383-1393 (2019).

[6] Yao, W. et al. Impact of Cone-Shape-Patterned Sapphire Substrate and Temperature on the Epitaxial Growth of p-GaN via MOCVD. *Phys. Status Solidi A* **216**, 1900026 (2019).

[7] Chen, Y. et al. Growth of 2D GaN single crystals on liquid metals. *J. Am. Chem. Soc.* **140,** 16392−16395 (2018).

[8] Syed, N. et al. Wafer-sized ultrathin gallium and indium nitride nanosheets through the ammonolysis of liquid metal derived oxides. *J. Am. Chem. Soc.* **141,** 104–108 (2019).

[9] Lee, L. Y. et al. Investigation of MOVPE-grown zincblende GaN nucleation layers on 3C-SiC/Si substrates. *J. Cryst. Growth* **524**, 125167 (2019).

[10] Kim, H. Vertical Schottky Contacts to Bulk GaN Single Crystals and Current Transport Mechanisms: A Review. *J. Electron. Mater.* **50**, 6688 (2021).

[11] Henning, A. et al. Aluminum Oxide at the Monolayer Limit via Oxidant-Free Plasma-Assisted Atomic Layer Deposition on GaN. *Adv. Funct. Mater.* **31**, 2101441 (2021).

[12] Rouf, P. et al. Hexacoordinated Gallium (III) Triazenide Precursor for Epitaxial Gallium Nitride by Atomic Layer Deposition. *Chem. Mater.* **33**, 3266-3275 (2021).

[13] Austin, A. J. et al. High-Temperature Atomic Layer Deposition of GaN on 1D Nanostructures. *Nanomaterials* **10**, 2434 (2020).

[14] Banerjee, S., Aarnink, A. A. I., Gravesteijn, D. J. & Kovalgin, A. Y. Thermal Atomic Layer Deposition of Polycrystalline Gallium Nitride. *J. Phys. Chem. C* **123**, 23214-23225 (2019).

[15] Liu, S. et al. Baking and plasma pretreatment of sapphire surfaces as a way to facilitate the epitaxial plasma-enhanced atomic layer deposition of GaN thin films. *Appl. Phys. Lett.* **116**, 211601 (2020).

[16] Song, Y. et al. Exploration of Monolayer MoS2 Template-Induced Growth of GaN Thin Films via Plasma-Enhanced Atomic Layer Deposition. *Cryst. Growth Des.* **21**, 1778-1785 (2021).

[17] Qiu, P. et al. Plasma-enhanced atomic layer deposition of gallium nitride thin films on fluorine-doped tin oxide glass substrate for future photovoltaic application. *Ceram. Int.* **46**, 5765-5772 (2020)).

[18] He, Y. F. et al. Growth of Gallium Nitride Films on Multilayer Graphene Template Using Plasma-Enhanced Atomic Layer Deposition. *Acta Metall Sin-ENGL*, **32**, 1530-1536 (2019).

[19] Wei, H. et al. Plasma-enhanced atomic-layer-deposited gallium nitride as an electron transport layer for planar perovskite solar cells. *J. Mater. Chem. A*, **7**, 25347–25354 (2019).

[20] Romo-García, F. et al. Gallium nitride thin films by microwave plasma-assisted ALD. *Opt. Mater. Express* **9**, 4187-4193 (2019).

[21] Gungor, N. & Alevli, M. Visible/infrared refractive index and phonon properties of GaN films grown on sapphire by hollow-cathode plasma-assisted atomic layer deposition. *J Vac Sci Technol A* **37**, 050901 (2019).

[22] Sadaf, S. M. & Tang, H. Mapping the growth of p-type GaN layer under Ga-rich and N-rich conditions at low to high temperatures by plasma-assisted molecular beam epitaxy. *Appl. Phys. Lett.* **117**, 254104 (2020).

[23] Mauraya, A. K. et al. Structural and ultraviolet photo-detection properties of laser molecular beam epitaxy grown GaN layers using solid GaN and liquid Ga targets. *Phys. Scr.* **96**, 085801 (2021).

[24] Dewan, S., Tomar, M., Tandon, R. P. & Gupta, V. In-situ and post deposition analysis of laser MBE deposited GaN films at varying nitrogen gas flow. *Vacuum* **164**, 72-76 (2019).

[25] Reilly, C. E. et al. Flow modulation metalorganic vapor phase epitaxy of GaN at temperatures below 600 ºC. *Semicond. Sci. Technol.* **35**, 095014 (2020).

[26] Liu, W. S., Chang, Y. L. & Chen, H. Y. Growth of GaN thin film on amorphous glass substrate by direct-current pulse sputtering deposition technique. *Coatings* **9**, 419 (2019).

[27] Mantarci, A. & Kundakci, M. Power-dependent physical properties of GaN thin films deposited on sapphire substrates by RF magnetron sputtering. *Bull. Mater. Sci.* **42**, 196 (2019).

[28] Mantarcı, A. A Role of RF Power in Growth and Characterization of RF Magnetron Sputtered GaN/glass Thin Film. *Emerg. Mater. Res.* **8**, 1-12 (2019).

[29] Mantarcı, A & Kundakçi, M. Physical properties of RF magnetron sputtered GaN/n-Si thin film: impacts of RF power. *Opt. Quant. Electron.* **51**, 81 (2019).